\documentclass[pra,twocolumn,floatfix]{revtex4-1}

\usepackage[utf8]{inputenc}
\usepackage[backref=none]{hyperref}
\hypersetup{
    colorlinks=true,          
    linkcolor=blue,           
    citecolor=blue,           
    urlcolor=green,           
}

\usepackage{amsmath}
\usepackage{amssymb}
\usepackage{multirow}
\usepackage{graphicx}
\graphicspath{ {images/} }
\usepackage{array}
\usepackage{upgreek}

\usepackage[switch]{lineno}

\makeatletter

\begin{document}

\title{A coherent superposition of Feshbach dimers and Efimov trimers.}

\author{Yaakov~Yudkin$^{1}$}
\author{Roy~Elbaz$^{1}$}
\author{P.~Giannakeas$^{2}$}
\author{Chris~H.~Greene$^{3}$}
\author{Lev~Khaykovich$^{1}$}

\affiliation{$^{1}$Department of Physics, QUEST Center and Institute of Nanotechnology and Advanced Materials, Bar-Ilan University, Ramat-Gan 5290002, Israel}
\affiliation{$^{2}$Max Planck Institute for the Physics of Complex Systems, N\"othnitzer Strasse 38, 01187 Dresden, Germany}
\affiliation{$^{3}$Department of Physics and Astronomy, Purdue University, West Lafayette, Indiana 47907, USA }

\date{\today}

\begin{abstract}
A powerful experimental technique to study Efimov physics at positive scattering lengths is demonstrated.
We use the Feshbach dimers as a local reference for Efimov trimers by creating a coherent superposition of both states.
Measurement of its coherent evolution provides information on the binding energy of the trimers with unprecedented precision and yields access to previously inaccessible parameters of the system such as the Efimov trimers' lifetime and the elastic processes between atoms and the constituents of the superposition state.
We develop a comprehensive data analysis suitable for noisy experimental data that confirms the trustworthiness of our demonstration. 
\end{abstract}

\maketitle

In few-body physics, the laws of quantum mechanics allow formation of peculiar loosely bound states~\citep{Efimov70}.
Being insensitive to the details of the short range inter-particle interactions, they display a variety of universal properties~\citep{Braaten&Hammer06}.
In recent years ultracold atoms have emerged as a main experimental platform to explore universality in few-body systems~\citep{Greene17,Naidon17,D'Incao18}.
In the two-body domain, weakly bound dimers are now routinely used for the characterization of Feshbach resonances~\citep{Chin10} and serve as the initial state for the production of ultra-cold molecules in their ro-vibrational ground state~\citep{Bohn17}.
In the three-body domain, the captivating subject of Efimov physics has been explored in a variety of atomic systems~\citep{Kraemer06,Wenz09,Zaccanti09,Gross09,Pollack09,Ferlaino09,Zenesini13,Wild12,Tung14,Pires14,Johansen17}.
But, interestingly, experimental techniques used in these explorations have been essentially limited to the study of inter-atomic inelastic processes, such as three-body recombination. 
Such an approach is best suited for the region of negative scattering lengths ($a<0$), where trimers can be associated from the three atom continuum. 
In contrast, for positive scattering lengths ($a>0$) the presence of dimers shifts the recombination related loss features into the atom-dimer continuum and Efimov resonances remain inaccessible for direct observation.

One of the central results of experimental research reveals an intriguing universality in the absolute position of the Efimov three-body resonance across diverse open-channel-dominated Feshbach resonances~\citep{Berninger11} in a variety of atomic species~\citep{Gross10,Wild12,Roy13}.
The Efimov resonance's position is determined by the three-body parameter (3BP) which has generally been accepted to be system dependent, i.e. sensitive to the short-range physics.
However, experimental results urged the theory to reinterpret the 3BP.
Indeed, it was shown that this universality stems from the fact that atoms interact through a van der Waals potential which suppresses the probability to find the particles at short distances from each other~\cite{Wang12,YWang12,Schmidt12,Naidon14,Blume15,Langmack18}. 
The break-down of this universality has been predicted to occur only near closed-channel-dominated Feshbach resonances~\citep{Petrov04,Gogolin08,Sorensen12} which was confirmed in a recent experiment~\citep{Johansen17}.

\begin{figure}
\centering
\includegraphics[width=1.\columnwidth]{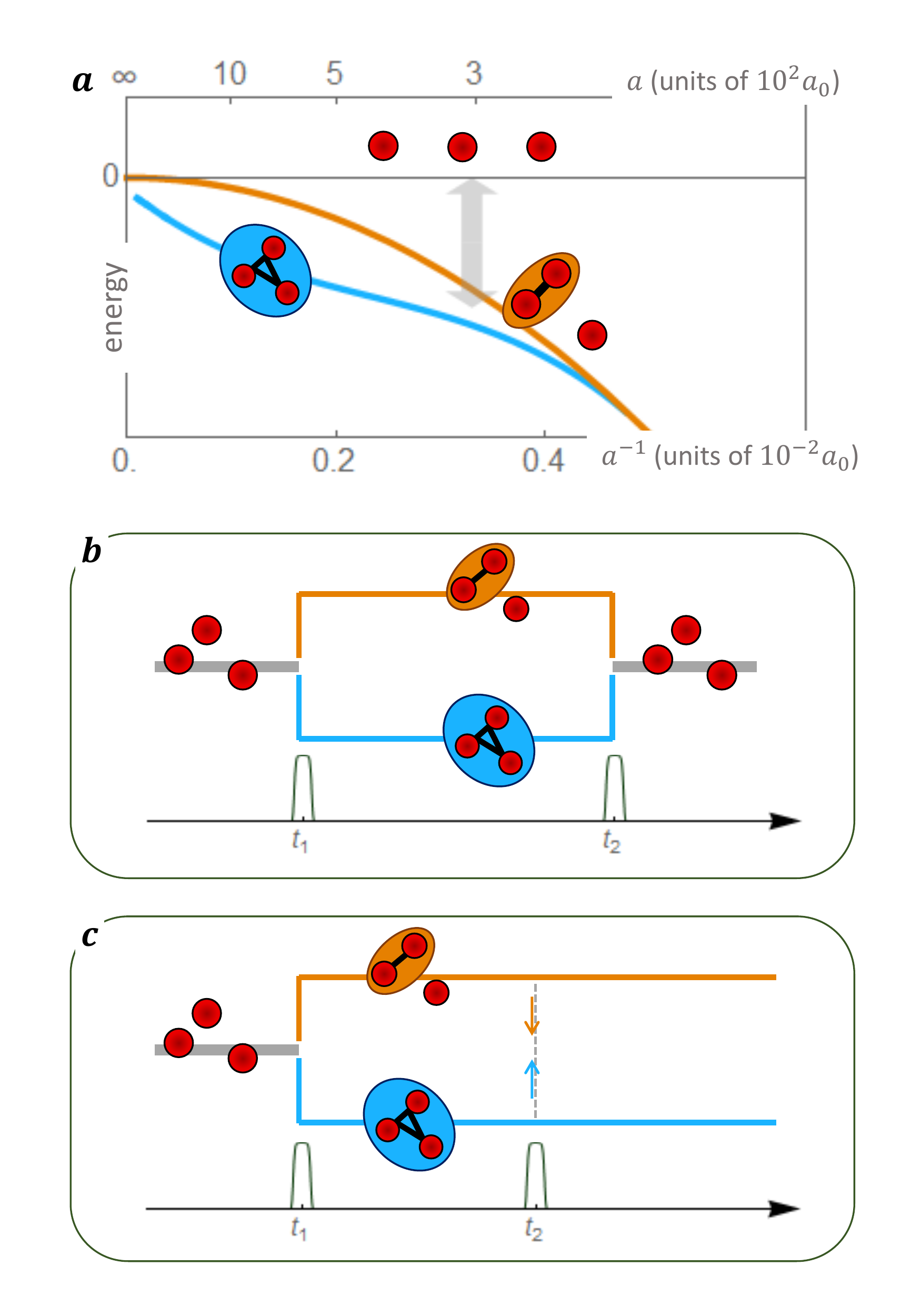}
\caption{\label{fig:interferometer}
\textbf{Illustration of the energy levels and of the interferometer.}
\textbf{a.}
The dimer (orange) and trimer (blue) energy levels (not to scale) are shown schematically as a function of the inverse scattering length.
The grey arrow indicates the parameter regime of our work and illustrates the effect of the modulation pulse.
\textbf{b,c.}
The two pulse sequence results in constructive (\textbf{b}) or  destructive (\textbf{c}) interference.
}
\end{figure}

The Efimov-van der Waals universality is well established for $a<0$.
However, for $a>0$ the situation remains obscure due to the lack of reliable experimental data.
In this range, Efimov trimers dissociate into an atom-dimer continuum [see Fig.~\ref{fig:interferometer}(a)] which is significantly more difficult to access experimentally. 
The way the dimers are created poses challenges to the preparation of an initial atom-dimer mixture with a significant population of dimers.
Among the various laser-cooled bosonic species only cesium allowed the preparation of a suitable atom-dimer mixture through a rather sophisticated protocol~\citep{Knoop09,Zenesini14}.
In other bosonic species Efimov resonances have been studied only indirectly by means of an avalanche mechanism~\citep{Zaccanti09,Pollack09,Dyke13,Machtey12,Machtey12a,Hu14} which is currently considered to be questionable~\citep{Langmack12,Zenesini14,Hu14}. 
From the theory side, the universality of the 3BP is predicted to weaken for  $a>0$~\citep{Kievsky13,Ji15,Giannakeas17,Mestrom17,Mestrom19}.
Moreover, it predicts that the first excited Efimov level avoids merging with the atom-dimer continuum due to various finite range effects~\citep{Mestrom17}.
Despite continuous theoretical interest, experimental progress is hindered due to inherent limitations of the currently available experimental techniques.

In a few previous experiments a limited range of the first excited Efimov energy level has been probed by RF association of distinguishable fermions~\citep{Lompe10RF,Nakajima11} and homonuclear bosons~\citep{Machtey12}.
The signature of production of Efimov trimers was revealed by loss resonances (inelastic collisions) which did not permit measuring their lifetime.
In addition, the finite resolution of the method prevented exploration of the vicinity of the atom-dimer resonance.
In a recent experiment with $^{85}$Rb the lifetime of the Efimov trimer has been directly measured for the first time~\citep{Klauss17}.
The applied method, however, prevented access to spectroscopic information on the Efimov energy level and required a rather significant initial population of trimers. 
 
This Letter demonstrates a new experimental approach to study the physics of Efimov trimers at positive scattering lengths.
As illustrated in Fig.~\ref{fig:interferometer}, we utilize a short and strong pulse of magnetic field modulation which is broad enough to create a coherent superposition state of Feshbach dimers and Efimov trimers.
After a variable time of its coherent evolution we apply a second pulse to observe the accumulated phase difference between the two constituents of the superposition state.
The resulting oscillations reveal the Efimov trimer energy level with nearly 10-fold improvement in precision and much higher resolution limit compared to the previously applied experimental method~\cite{Machtey12}.
Even more importantly, we observe the decay of the coherent oscillations which can be related to different decoherence mechanisms such as the trimers' lifetime and the elastic atom-dimer and/or atom-trimer collision rates.
Note, finally, that we benefit from the four-fold interferometric enhancement of the signal and demonstrate high sensitivity in probing a small population of trimers.

The experiment is performed on $^7$Li atoms, polarized in the $|F=1, m_{F}=0\rangle$ state and evaporatively cooled to a temperature of $T\approx1.5\;\mu$K in a crossed-beam optical trap in the vicinity of a Feshbach resonance~\cite{supMat}.
The magnetic field bias is set to $880.25$~G which corresponds to a scattering length of $\sim300a_0$ and a Feshbach dimer binding energy of $E_d=-h\times6$~MHz [see Fig.~\ref{fig:interferometer}(a)]~\cite{Gross11}.
According to an earlier study described in Ref.~\cite{Machtey12}, the first excited Efimov trimer energy level is predicted to be just $\sim100$~kHz below $E_d$, i.e. ${E_t-E_d\approx -h\times100}$ kHz, where $E_t$ is the energy of the trimer state.
In Ref.~\cite{Machtey12}, $E_t-E_d$ was measured in the region of $0.5$~MHz $<E_d/h<4$~MHz by means of loss spectroscopy.
The frequency dependent magnetic field modulation was applied for tens to hundreds of ms, and the induced atom loss resonances were related to the positions of dimer and trimer energy levels.
The accessible region of the trimer's binding energies was constrained by the finite resolution limit to be $\gtrsim 110$~kHz.
Thus, our current measurement probes the region which was out of reach for the previous experimental technique.

\begin{figure*}
\centering
\includegraphics[width=1.\textwidth]{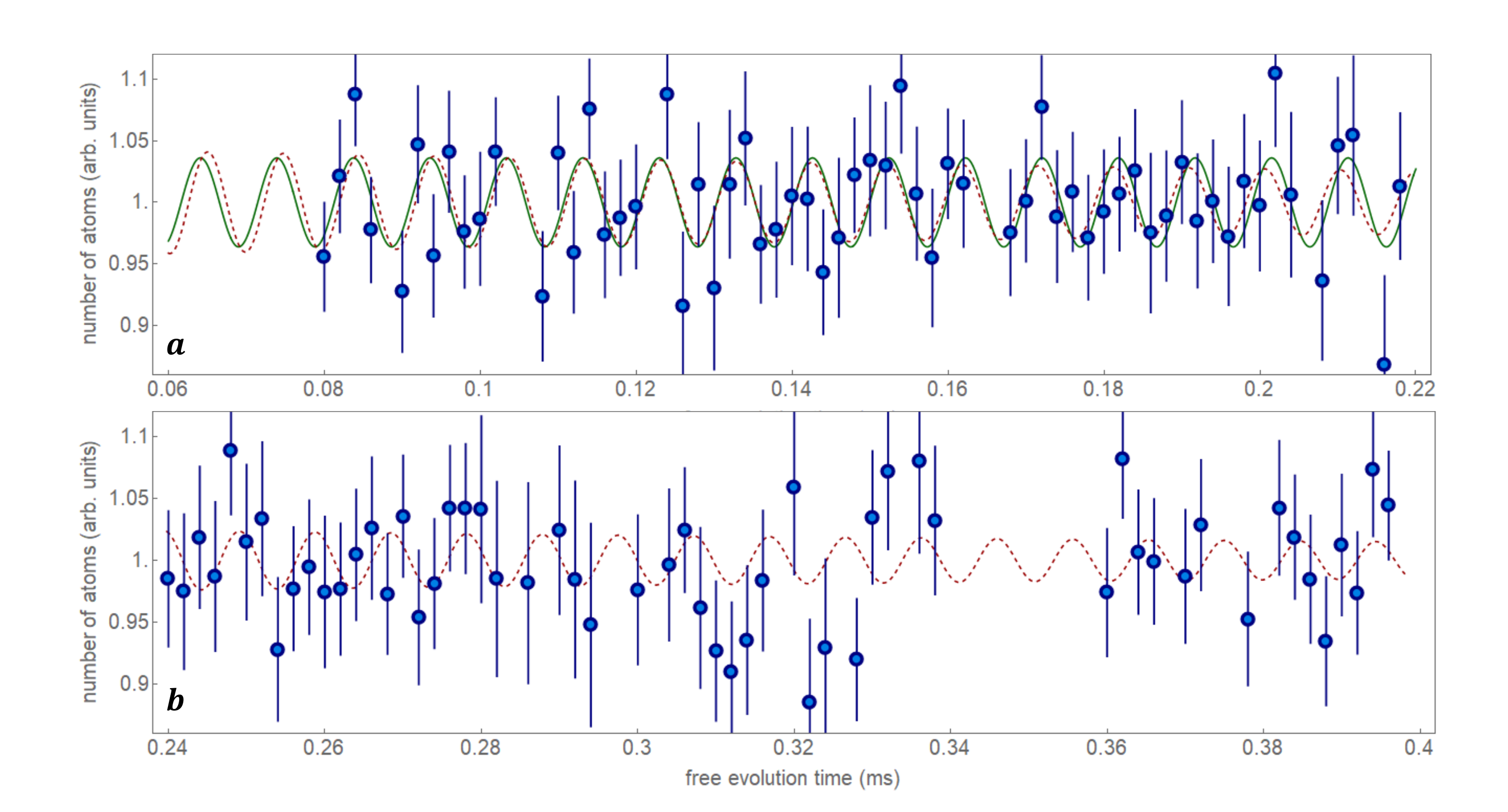}
\caption{\label{fig:numbVStime}
\textbf{Observation of oscillations.}
The number of free atoms $N$ after the interferometer sequence is measured as a function of the free evolution time $t$.
Each data point is the average of $2-8$ measurements and the error bar is $1\sigma$ of the mean error.
\textbf{a.}
Data for $t\in[80,220]\;\mu$s.
The green line is the best fit to a cosine.
\textbf{b.}
Data for $t\in[240,400]\;\mu$s.
The red dashed line in \textbf{a.} and \textbf{b.} is (one and the same) best fit to a decaying cosine.
}
\end{figure*}

The concept of the interferometer relies on a clear separation of energy scales in the system: $E_d \gg E_t-E_d > k_B T$.
The first step of the interferometer, shown in Fig.~\ref{fig:interferometer}(b,c), is the beam-splitter at time $t_1$.
The bias field is modulated at $\nu_{m}=6$ MHz for a FWHM duration of $\tau_{m} = 10\;\mu$s by a single auxiliary coil.
The modulation amplitude at the atom position is $b\approx1.5$~G~\cite{supMat}.
The Fourier transform limited bandwidth of the pulse is $100$ kHz at FWHM which allows us to address both (dimer and trimer) energy levels simultaneously while covering the full thermal distribution of the free-atom continuum ($\sim$30~kHz for $T\approx1.5\;\mu$K).
The pulse projects the three-atom continuum to a coherent superposition state of a dimer + free atom and a trimer, denoted hereafter as DITRIS (DImer-TRImer Superposition) state.
The system then evolves freely for a variable time $t\gg\tau_{m}$ during which the two constituents of the DITRIS accumulate a relative phase difference of $\phi(t) = (E_t-E_d)t/\hbar$, assuming that the energy of the free atom in the dimer + atom pathway is negligible (see the discussion of the results below).
At time $t_2=t_1+t$  an identical modulation pulse projects the two paths back to free atoms and serves as an output port of the interferometer.
When $\phi(t)=2\pi n$, where $n \in\mathbb{Z}$, constructive interference between the two paths projects the three atoms into the three-atom continuum as shown in Fig.~\ref{fig:interferometer}(b).
In contrast,  Fig.~\ref{fig:interferometer}(c) represents the case where $\phi(t)=\pi(2n+1)$ when destructive interference suppresses dissociation of the bound states.
This produces a time dependent periodic variation in the number of free atoms with a peak-to-peak amplitude proportional to $N_{D}$, where $N_D$ is the number of DITRIS states produced by the first pulse.
This two-path interferometer picture neglects the contribution of the third path where the three atoms remain in the three-atom continuum.
However, due to our experimental conditions ($E_d \gg E_t-E_d$) this channel contributes oscillations at $\sim E_d/h$ which are averaged to zero in the range of interest, namely $(E_t - E_d)/h$~\citep{supMat}.

The results of our interferometer at the output port are shown in Fig.~\ref{fig:numbVStime}, where we measure the number of free atoms, $N(t)$, as a function of the free evolution time $t$.
Each point represents the mean of 2 - 8 individual measurements. 
It is evident from the data that the signal-to-noise ratio (SNR) is small. 
We, hence, begin the analysis with the data for short evolution times [$80$ $\mu$s $<t<220$ $\mu$s; Fig.~\ref{fig:numbVStime}(a)], where the oscillations can be visually appreciated.
Assuming constant amplitude oscillations, we apply three different analyses:  (i) a fast Fourier transform (FFT), and (ii) a two- and (iii) a three-parameter fit to the cosine function:
\begin{equation}
\frac{N(t)}{N_0} = 1 + A\cos\left(2\pi\nu t+\varphi\right),
\label{eq:sine}
\end{equation}
where $N_0$ is the initial number of atoms in the trap.
In (ii), the two fitting parameters are the amplitude $A$ and the phase $\varphi$, while in (iii) the frequency $\nu$ becomes the third fitting parameter.

\begin{figure}
\centering
\includegraphics[width=1.\columnwidth]{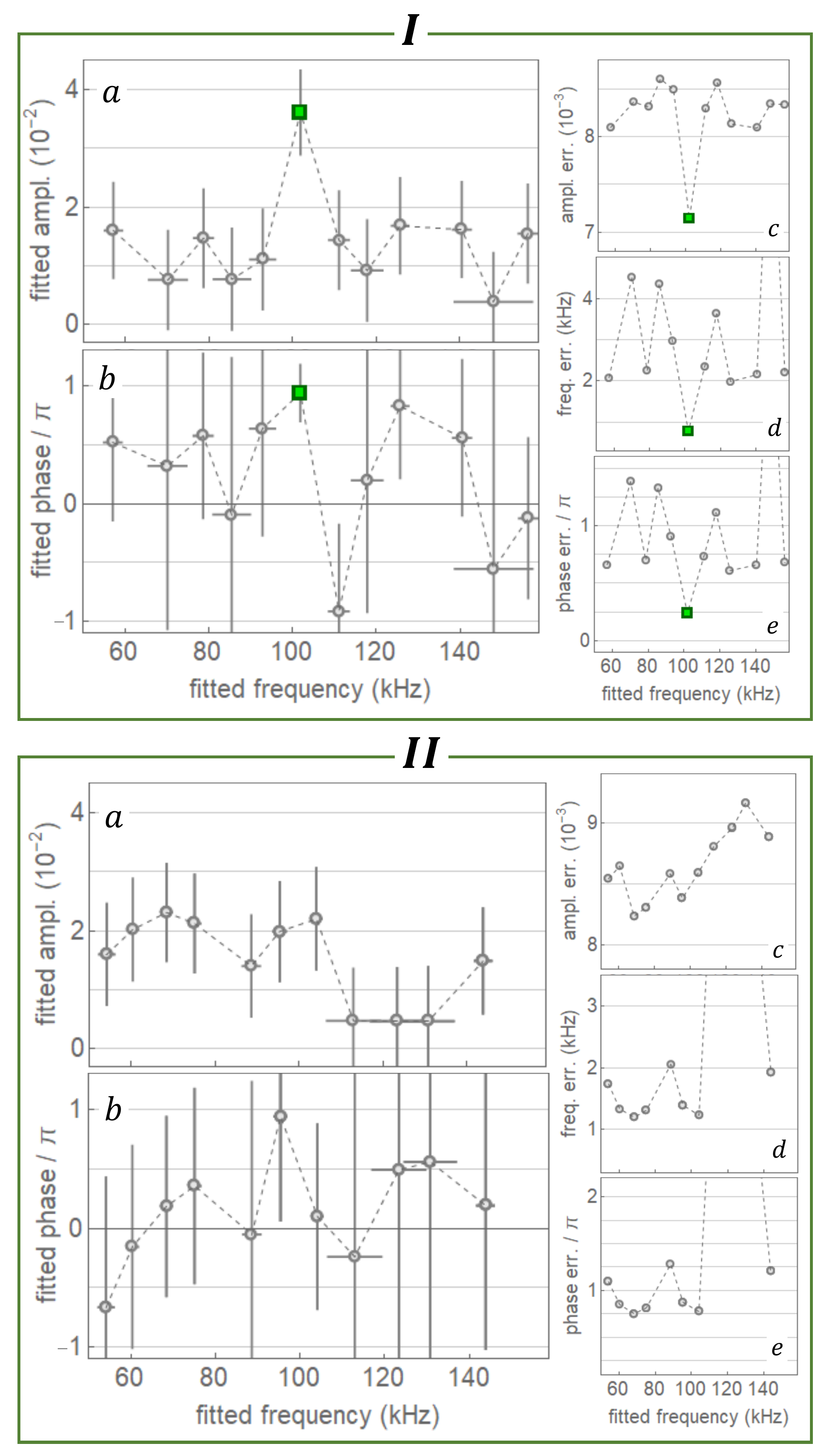}
\caption{\label{fig:3par_osc}
\textbf{Data analysis using three free parameters.}
\textbf{I.}
Using the data of Fig.~\ref{fig:numbVStime}(a) and Eq.~(\ref{eq:sine}), the results of a three-parameter fit for different initial parameter values are shown: (\textbf{a}) $A(\nu)$ and (\textbf{b}) $\varphi(\nu)$.
The error bars are $1\sigma$ fitting uncertainties and they are also depicted in (\textbf{c}), (\textbf{d}) and (\textbf{e}) as a function of $\nu$.
The parameters indicated by the green square are used for the green fit in Fig.~\ref{fig:numbVStime}(a).
\textbf{II.}
Same as (\textbf{I}) for the data of Fig.~\ref{fig:numbVStime}(b).
}
\end{figure}

We verify that in case of small SNR, the most stringent test for the claim that an oscillation frequency has been observed in the experiment is provided by the three-parameter fit, and, therefore, only this analysis is discussed below~\citep{supMat}.
The fitting algorithm is applied to the experimental results using a variety of initial conditions. 
In particular, the initial frequency is scanned very densely in the relevant frequency range.
However, the fitting algorithm converges only to a limited number of frequencies, all of which are shown in Fig.~\ref{fig:3par_osc}(I).
Among all the converged frequencies, only $\nu^\star=102.9(8)$ kHz has a distinguishable amplitude $A^\star=0.036(7)$, while all other $A(\nu)$ are smaller and similar to each other (see Fig.~\ref{fig:3par_osc}(I)(a)).
This provides the first evidence that a single, dominant frequency can be indicated in the data.
In fact, a naturally defined SNR $=A^\star/\bar{A}$, where $\bar{A}$ is the mean value of $A(\nu)$ excluding $A^{*}$, is found to be SNR $=2.79(55)$.
The second evidence for the presence of a dominant frequency is the coincidence of global minima in errors obtained for all three fitting parameters at $\nu^\star$ [see Fig.~\ref{fig:3par_osc}(I)(c,d,e)].

The reported small SNR poses an obvious question: what is the likelihood to observe a similar peak with the same frequency analysis for a randomly generated noisy data?
This is exactly the question where the full strength of the three-parameter fit method is revealed~\citep{supMat}.
We perform a likelihood analysis with $1000$ fake sequences of $N(t)$ with the same sampling frequency and length as in Fig.~\ref{fig:numbVStime}(a) generated from a random Gaussian distribution with the same standard deviation as in our experimental data.
We then apply our frequency analysis for each sequence and look for events with a SNR larger than $1\sigma$ below the experimental SNR and a central frequency within the expected interval of $[90,110]$ kHz.
The likelihood analysis results in zero such events.
Only when the SNR-threshold is lowered to $2.1$ ($1.25\sigma$ below the experimental SNR) the first false positive occurs.
Hence the probability of our result being wrong is $0.21\times0.001=2\times10^{-4}$ ($0.21$ is the probability of getting a result $\geq1.25\sigma$).
This sufficiently negligible probability together with the fact that no oscillations were detected when only a single pulse was applied~\citep{supMat} allows us to fully trust our results.

\begin{figure}
\centering
\includegraphics[width=1.\columnwidth]{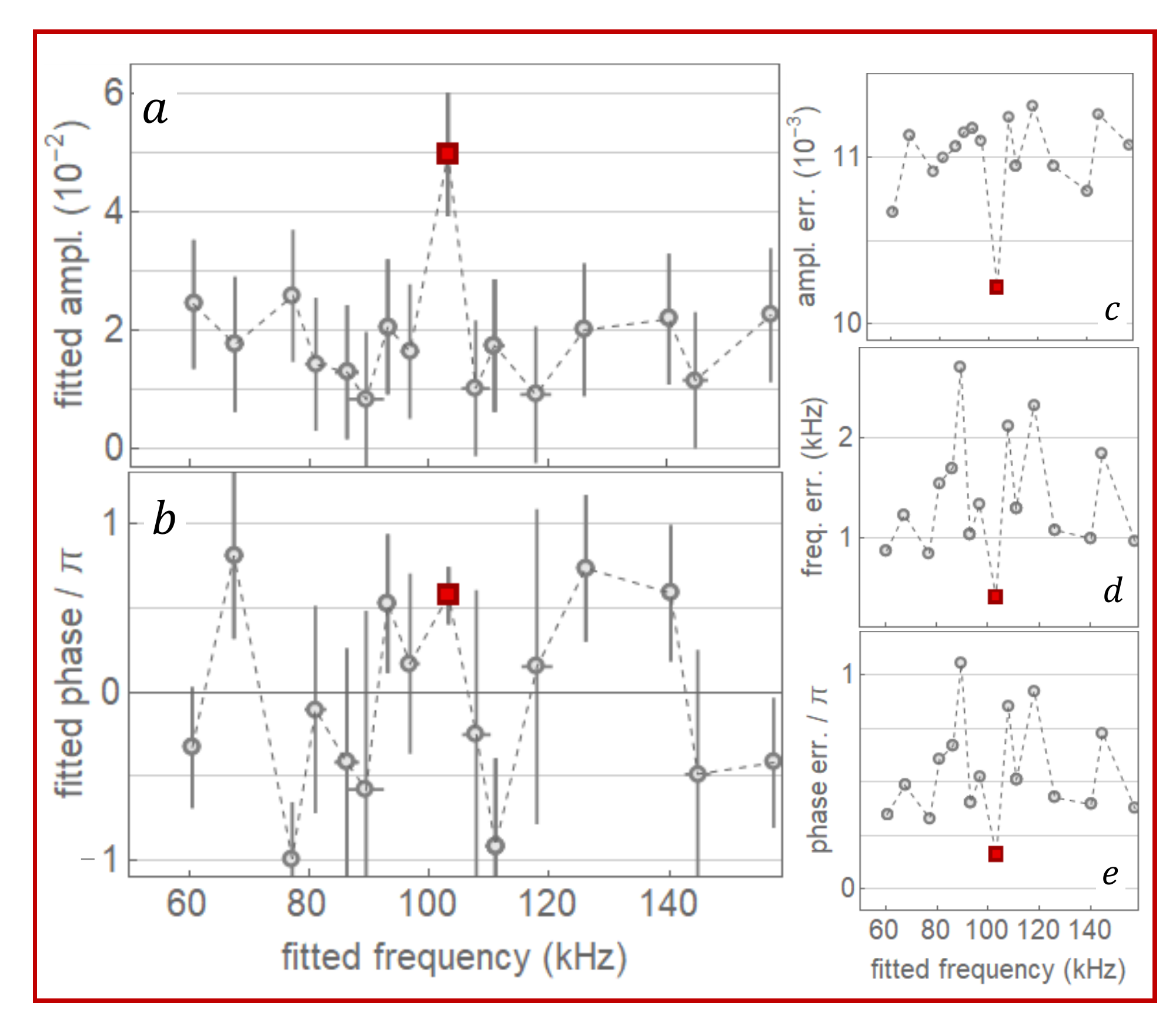}
\caption{\label{fig:3par_dec}
\textbf{Data analysis using three free parameters and a decaying cosine.}
Same as Fig.~\ref{fig:3par_osc}, but for the entire data set [Fig.~\ref{fig:numbVStime}(a+b)] and using Eq.~(\ref{eq:decsine}) for the fit.
The parameters indicated by the red square are used for the red (dashed) fit in Fig.~\ref{fig:numbVStime}.}
\end{figure}

To see the duration of the coherent oscillations, the experiment is repeated for longer free evolution times.
In Fig.~\ref{fig:numbVStime}(b) the data points for $t$ between $240\;\mu$s and $400\;\mu$s are shown, and the corresponding three-parameter fit analysis is in Fig.~\ref{fig:3par_osc}(II).
Since we observe no dominant contribution at any frequency we conclude that the oscillations are below the detection limit and, thus, we detect the decay of the signal.
Next, the entire data set ($t\in\left[80,400\right]$) is analyzed by fitting it to a damped cosine curve:
\begin{equation}
\frac{N(t)}{N_0}=1 + Ae^{-t/\tau}\cos\left(t \sqrt{(2\pi\nu)^2+\tau^{-2}}+\varphi\right).
\label{eq:decsine}
\end{equation}
This analysis is performed in two steps.
First, a fit with four free parameters ($A$, $\tau$, $\nu$ and $\varphi$) is executed in the vicinity of the earlier derived values of some of them ($A^\star,\,\nu^\star$ and $\varphi^\star$).
The fit converges and yields $\tau=331(297)\;\mu$s which allows us to put an upper bound of $\sim 630\;\mu$s for the detected decay time.
The three-parameter fit analysis is then repeated for the entire data set using Eq.~(\ref{eq:decsine}) and keeping $\tau=331\;\mu$s fixed.
The result, shown in Fig.~\ref{fig:3par_dec}, agrees with the previous one [Fig.~\ref{fig:3par_osc}(I)] for the central frequency $\nu^\star$ (within the errors) and has a slightly reduced SNR $=2.41(49)$.
The phase of the oscillations at $\nu^\star$ is measured to be $\varphi^\star=0.57(16)\pi$, which is consistent with $N(t=0)=N_0$.

Extrapolating the damped oscillations to $t=0$ we obtain an initial peak-to-peak signal of ${2A=0.10(1)\times N_0\approx 3,000}$ atoms, where ${N_0\approx3\times 10^{4}}$ is the initial number of atoms before the first pulse is applied.
This implies that $\sim3,000/4=750$ atoms participate in a DITRIS corresponding to $N_D\approx750/3\approx250$~\cite{supMat}.
Our sensitivity is thus limited to $\sim100$ DITRIS states (according to the SNR) and we fully benefit from the interferometric enhancement of the signal to obtain this level of sensitivity.
For our experimental parameters, a trimer-excluding theory predicts the conversion of $\sim 0.018\times N_0 \approx 540$ atoms into $\sim 270$ dimers after the first pulse~\citep{Giannakeas18}.
This provides an upper bound for DITRIS states that can be created in our system in agreement with the observed results.

We now discuss the decay of the coherent oscillations, which can be caused by different mechanisms.
The first is related to the finite thermal energy of the "spectator" atom in dimer + atom path of the interferometer. 
However, this scenario, if it were relevant, would cause a significantly faster decay as the thermal energy has a $\sim 30$~kHz bandwidth.
Therefore, experimental results signify a certain narrowing mechanism which currently remains unclear.
This question deserves special attention, which we plan to provide in a future combined experimental and theoretical effort.
On the other hand, this result opens up the opportunity to study other sources of decoherence.

The other two mechanisms indicate access to new observables, that were inaccessible in previous experiments.
In the first, molecules are lost due to the finite lifetime of the trimers ($T_1$ time).
As trimers naturally have a shorter lifetime than the dimers, the contribution of the finite lifetime of the dimers can be safely excluded from consideration.
In the second mechanism ($T_2$ time) coherence is lost due to elastic collisions between the constituents of the DITRIS states and the free atoms.
Our reported result sets an upper bound on both times.
Future experiments should be able to determine which of the above is the dominant mechanism by analyzing the density dependence of the decoherence time $\tau$.
The $T_1$ time is density independent while the $T_2$ time decreases for higher densities.

The observed decoherence is probably caused by the finite lifetime of the trimers (which seems to be longer than the previously observed lifetime in a gas of $^{85}$Rb atoms~\citep{Klauss17}).
However, closer to the Efimov resonance the atom-dimer elastic collisional cross-section increases by several orders of magnitude and might become the dominant mechanism for decoherence.
Future experiments should allow investigation of this aspect of Efimov physics which is currently totally absent from the list of available experimental observables.

In conclusion, we demonstrate the creation of a superposition state of Efimov trimers and Feshbach dimers.
The newly developed data analysis, consistency in the fitting parameters and vanishingly small level of failure in likelihood analysis prove that the interference signal between the constituents of the superposition state that we observe in the experiment is real.
We thus measure the energy of the trimer bound state with high precision and give an upper bound for decoherence processes.

An interesting question for future research is the identification of the exact mechanism responsible for this decoherence.
Another extension of our work consists of repeating the interferometer sequence at different values of the bias field (different scattering length) to explore the elastic collisions channel and to identify the position of the Efimov resonance, and whether it even exists.
Applying the interferometer to other accessible sub-levels of lithium atoms and other atomic species should clarify and deepen our understanding of finite range corrections to the Efimov physics at positive scattering lengths. 
Finally, we speculate that our interferometer might be extended to study tetramers and possibly higher multimers for which very little experimental results are available.

{\bf Acknowledgements}
Y.Y. and L.K. acknowledge fruitful discussions with N. Navon.
This research was supported in part by the Israel Science Foundation (Grant No. 1340/16), the United States-Israel Binational Science Foundation (BSF, Grant No. 2012504) and by NSF Grant No. PHY-1607180.
The numerical calculations have been performed using NSF XSEDE Resource Allocation No. TG-PHY150003.


\newpage

\setcounter{figure}{0}
\renewcommand\thefigure{S\arabic{figure}}   

{\huge Supplementary Material}

\section{Experimental details}
\subsection{Sample preparation} 
A far off-resonance (wavelength $1080$ nm) crossed-beam optical trap (OT) is loaded directly from a compressed magneto-optical trap.
By ramping down the intensity of the OT the atomic cloud is evaporatively cooled in the vicinity of a narrow Feshbach resonance~[47] and the magnetic field is then ramped up to the vicinity of a broad Feshbach resonance located at $893.7$ G.
In the final state the trapping frequencies of the radially symmetric trap are $\left(\omega_{r},\omega_{z}\right)\approx2\pi\times\left(1500,60\right)$ Hz in the radial and axial directions respectively.
The atoms have a peak density of $n_0\approx5\times10^{12}$ cm$^{-3}$, a temperature of $T\approx1.5\;\mu$K and a s-wave scattering length of  $a\approx300a_0$, corresponding to a bias field of $880.25\;$G.
The interferometer experiment is performed at this point.

\subsection{Setting the magnetic field}
The bias field $\bar{B}$ and modulation frequency $\nu_m$ have to be such that the dimer/trimer production is maximal.
We set the modulation frequency to $\nu_m=6$ MHz for all the data presented in this letter (see next subsection) and find the resonance of $\bar{B}$ spectroscopically.
To this end we use a $1$~ms long pulse and measure the number of remaining atoms as a function of $\bar{B}$.
At a certain bias magnetic field a loss of $\sim50\%$ of the atoms is observed caused by inelastic collisions between the associated molecules and the remaining atoms.
We associate this feature with the field where the maximal production efficiency of molecules is obtained.
We observe slight variations of the field on the pulse intensity~[48] but not on the pulse duration.
In virtue of their close vicinity, $E_d$ and $E_t$ are unresolved in this experiment~[34].

\subsection{Field modulation}
The essence of our experiment is the short ($10\;\mu$s) and strong ($1.5$~G) modulation of the bias field at high modulation frequency which serves as beam-splitters of the interferometer.
This pulse is generated by a home-built resonance LC-circuit, where the inductance (L) is provided by the auxiliary coil itself.
An arbitrary function generator creates a square pulse with a single polarity.
It is amplified by a digital amplifier (peak current $1.7$ A) and injected into the LC-circuit which serves as a filter for the resonance frequency.
The result is a high amplitude, sinusoidal modulation of the magnetic field.
The rise time of the modulation amplitude is 4$\mu$s and the FWHM of the pulse is 10$\mu$s.

\subsection{Data acquisition}
After the second pulse of the interferometer experiment, the atoms are counted following a $350\;\mu$s time-of-flight (TOF) by absorption imaging, whose frequency is tuned such that only free atoms are detected (dimers and trimers are out of resonance with the detection light).
Hence the interference is seen by plotting the number of atoms $N$ versus free evolution time $t$.
To obtain the data we run many, partially overlapping sequences of $40\;\mu$s (e.g. $t\in\left[120,160\right]\;\mu$s) and $60\;\mu$s (e.g. $t\in\left[80,140\right]\;\mu$s) making sure that the system remains stabilized in-between sequences.
In each sequence, the experiment is repeated several times and averaged for each $t$.
We observe a slight change in the mean atom number (offset of the oscillations) from sequence to sequence.
Thus, each sequence is normalized to its offset.
The sequences are then stitched together, where the mean value is taken for overlapping regions.

We have verified that no oscillations are detectable when using solely one pulse (on a $80\;\mu$s long sequence, $2-4$ measurements per point) and when the magnetic field is off-resonance ($40\;\mu$s sequence) for short evolution times.
We stress again, that in the long sequence for long evolution times [Fig.~\ref{fig:numbVStime}(b) of main text] also no oscillations are detectable.
In addition, at very long evolution times ($400\leq t\leq1000\;\mu$s), which we probed in several $40\;\mu$s intervals, an oscillating signal also remains undetectable.

\section{Interferometer considerations}
\begin{figure}
\centering
\includegraphics[width=1.\columnwidth]{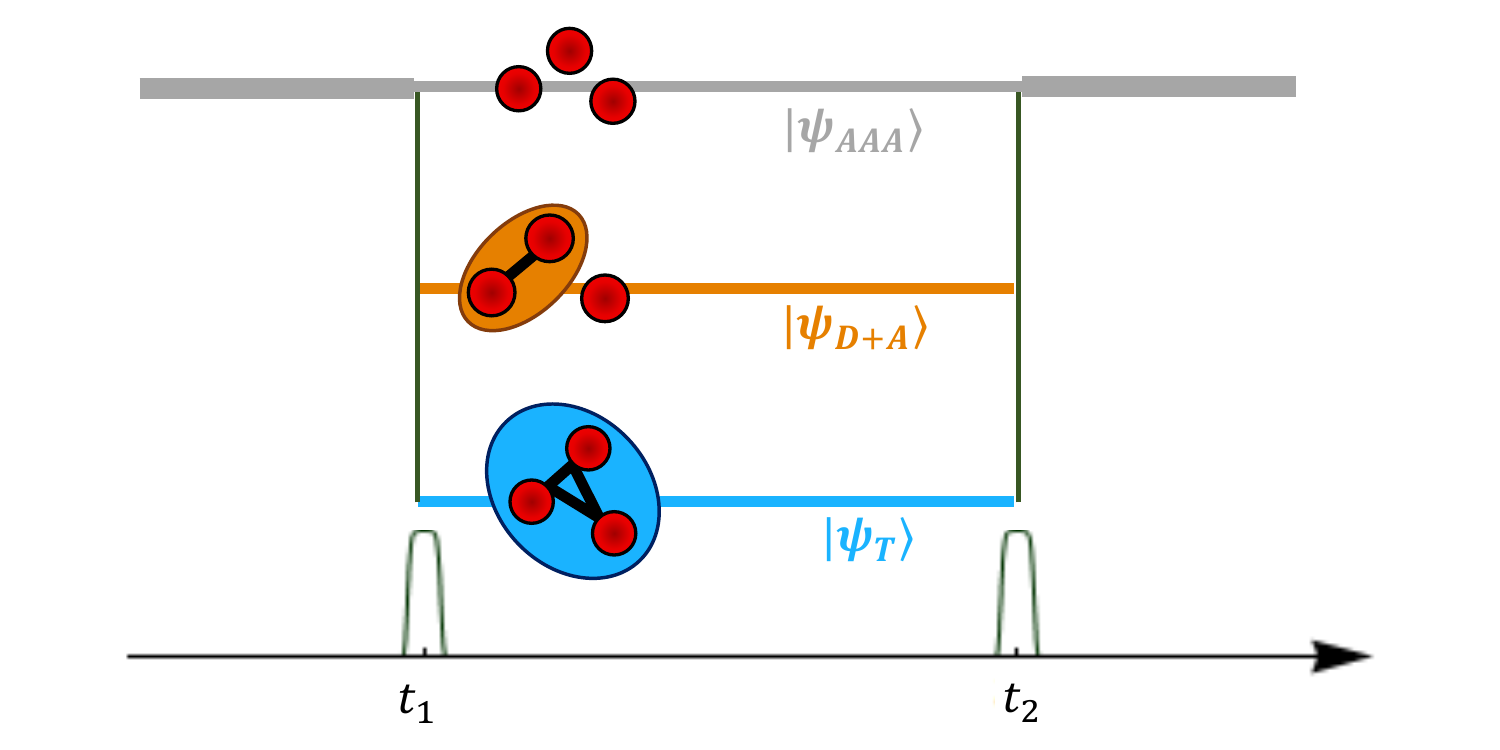}
\caption{\label{fig:interferometerWF}
\textbf{Illustration of the 3-particle wave function.}
}
\end{figure}

Here we refine the meaning of the DITRIS (see main text) by deriving its wave function based on the three-atom Hilbert space illustrated in Fig.~\ref{fig:interferometerWF}.

Initially the three atoms are free: $|\psi_{i}\rangle=|\psi_{AAA}\rangle$.
The first pulse (at time $t_1=0$) projects this state into a superposition of free atoms ($|\psi_{AAA}\rangle$), a dimer + atom ($|\psi_{D+A}\rangle$) and a trimer ($|\psi_{T}\rangle$):
\begin{equation}
|\psi(t_1)\rangle=A|\psi_{AAA}\rangle+B|\psi_{D+A}\rangle+C|\psi_{T}\rangle,
\end{equation}
where $A,B,C$ are the probability amplitudes for the conversion.
We neglect the contribution of dimers which are not in the superposition state and assume $B\approx C$ which, in fact, is the DITRIS association amplitude.
During the free evolution time $t=t_2-t_1$ the $|\psi_{D+A}\rangle$ and the $|\psi_{T}\rangle$ terms gain relative phases of $e^{-i\omega_{DA} t}$ and $e^{-i\omega_T t}$ respectively.
The second pulse (at time $t_2$) projects the two bound states back into the free atom continuum, i.e. $|\psi_{D+A}\rangle\rightarrow|\psi_{AAA}\rangle$ and $|\psi_{T}\rangle\rightarrow|\psi_{AAA}\rangle$, such that after the pulse the three particle system is described by
\begin{equation}
|\psi_f\rangle=\left(A+Be^{-i\omega_{DA} t}+Be^{-i\omega_T t}\right)|\psi_{AAA}\rangle.
\end{equation}
The probability of detecting all three atoms is given by the square of the coefficient.
Extracting a common phase factor of $e^{-i\omega_{DA} t}$ and keeping among the interference terms only those with slow varying phase $e^{-i(\omega_T-\omega_{DA}) t}$ we get:
\begin{equation}
P=A^2+2B^2\left[1+\cos(\omega t)\right],
\end{equation}
where $\omega = \omega_T-\omega_{DA}$.
The first term is the offset (or background) of the oscillation signal.
The second term, which has a peak-to-peak modulation of $4B^2$, is proportional to the detected oscillation signal and demonstrates the factor $4$ enhancement of the DITRIS signal exploited in the experiment.
To estimate the number of DITRIS states associated in the experiment, we must divide the detected peak-to-peak modulation in number of atoms by $4\times3=12$, because, in addition to the interferometric enhancement, there are three atoms in each DITRIS state.
As stated in the main text, we find that $3000/12\approx250$ such superposition states are created.

\section{Data analysis}
\subsection{Analysis (i): FFT}
\begin{figure}
\centering
\includegraphics[width=0.7\columnwidth]{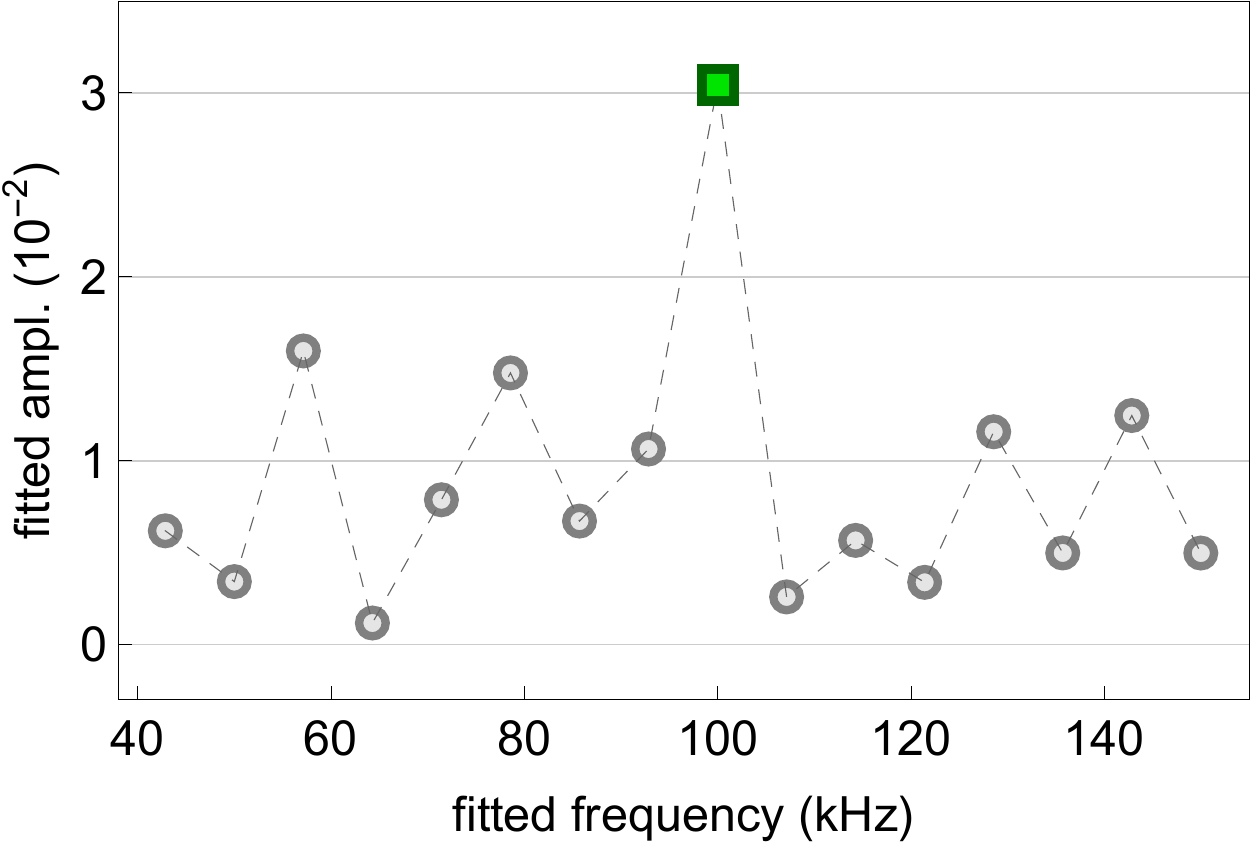}
\caption{\label{fig:FFT}
\textbf{FFT of data presented in Fig. 2(a) of the main text.}}
\end{figure}

In addition to the three-parameter fitting procedure described in the main text, the time-domain oscillation signal is analyzed using a numerical fast-Fourier-transform (FFT) and is shown to have the same qualitative behavior.
In Fig.~\ref{fig:FFT} the FFT of the interval $t\in\left[80,220\right]\;\mu$s [the data in Fig.~\ref{fig:numbVStime}(a) of the main text] is shown.
The $\delta$-like feature obtained has a signal-to-noise ratio of SNR $=3.57$ and is centered at $\nu^\star=100$ kHz.
Since FFT is model-independent and since the peak appears at the expected frequency value these observations are a convincing indication of a single dominant frequency contribution.
Due to our limited sample size ($N_s=70$), the width of each bin is $\nu_s/N_s\approx7.1$ kHz, where $\nu_s=500$ kHz is the sampling frequency.
The fitting procedures are therefore more accurate than the FFT in the determination of the oscillation frequency.

\subsection{Analysis (ii): two-parameters fit}
\begin{figure}
\centering
\includegraphics[width=1.\columnwidth]{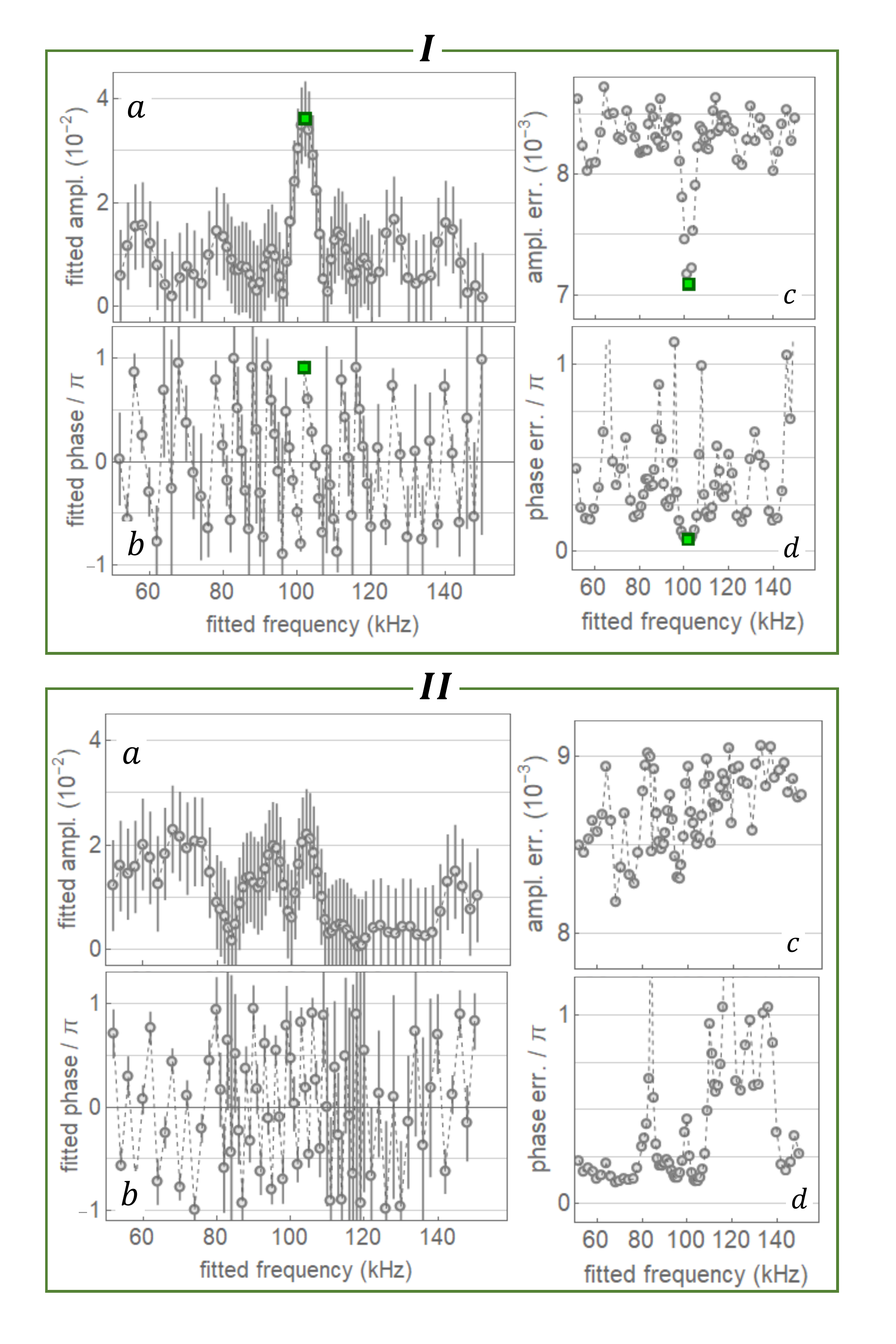}
\caption{\label{fig:2par_osc}
\textbf{Data analysis using two free parameters and a constant-amplitude cosine.}
\textbf{I.} For the data in Fig.~\ref{fig:numbVStime}(a) of the main text, the fitted amplitude $A(\nu)$ (\textbf{a}) and phase $\varphi(\nu)$ (\textbf{b}) are shown as a function of the fixed frequency.
Subplots (\textbf{c}) and (\textbf{d}) are their respective $1\sigma$ fitting error.
The best fit is indicated by the green square.
\textbf{II.}
Same as (\textbf{I}) for the data in Fig.~\ref{fig:numbVStime}(b) of the main text.
}
\end{figure}

\begin{figure}
\centering
\includegraphics[width=1.\columnwidth]{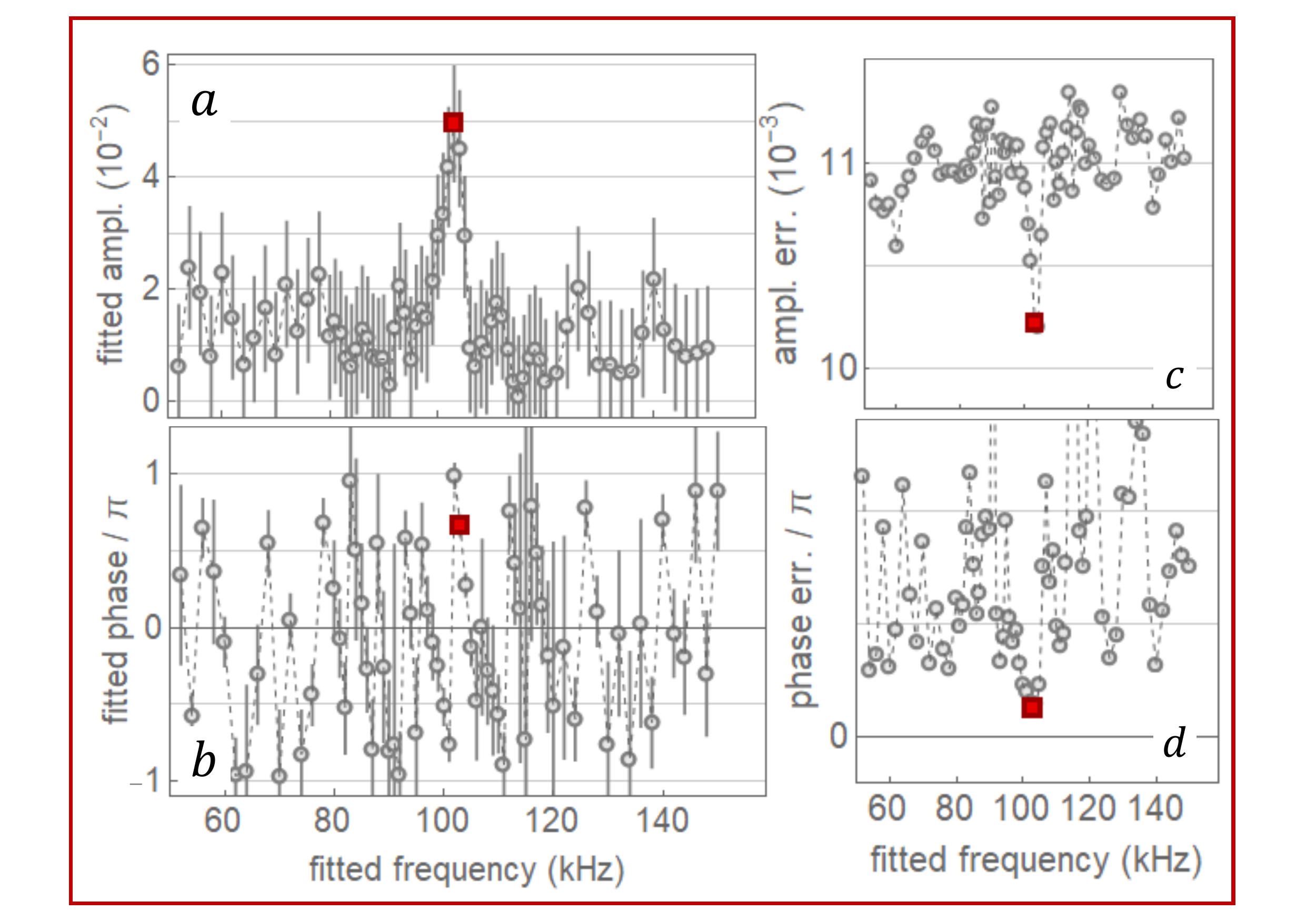}
\caption{\label{fig:2par_dec}
\textbf{Data analysis using two free parameters and a decaying cosine.}
Same as Fig.~\ref{fig:2par_osc}, but for the entire data set [Fig.~\ref{fig:numbVStime}(a+b) in main text] and using the decaying cosine function.
The best fit is indicated by the red square.
}
\end{figure}

Here, a second alternative to the three-parameters fit is considered.
The cosine function in Eq.~(\ref{eq:sine}) of the main text is fitted to the experimental data while holding the frequency $\nu$ fixed and using the amplitude $A$ and the phase $\varphi$ as fitting parameters.
This procedure is repeated for various values of $\nu\in[50,150]$ kHz.
In Fig.~\ref{fig:2par_osc}(I) the fitted amplitude (a) and phase (b) are plotted versus $\nu$ together with the $1\sigma$ fitting error (c,d).
As the frequency is fixed no fitting error is associated with it.
Then, the two-parameter fit is applied to the second time interval [Fig.~\ref{fig:2par_osc}(II)] and, using Eq.~(\ref{eq:decsine}) of the main text with fixed $\tau=331\;\mu$s, to the entire data set (Fig.~\ref{fig:2par_dec}).

Table.~\ref{tb:frequency} summarizes the results of all three methods.

\begin{table}
\centering
\begin{tabular}{c l|c c|c c|c c|c c}
& & \multicolumn{4}{c|}{constant cosine} & \multicolumn{4}{c}{decaying cosine} \\
\cline{3-10}
& & $\nu$ & $\Delta\nu$ & $\text{SNR}$ & $\Delta(\text{SNR})$ & $\nu$ & $\Delta\nu$ & $\text{SNR}$ & $\Delta(\text{SNR})$ \\
\hline
(i) & FFT & $100$ & $7.1$ & $3.57$ & - & - & - & - & - \\
(ii) & 2 par & $102$ & $1$ & $3.43$ & $0.67$ & $103$ & $1$ & $3.66$ & $0.75$ \\
(iii) & 3 par & $101.9$ & $0.8$ & $2.79$ & $0.55$ & $103.2$ & $0.4$ & $2.41$ & $0.49$
\end{tabular}
\caption{\label{tb:frequency}
\textbf{Observed frequency $\nu$ [kHz] and SNR for the various analyses.}}
\end{table}

\subsection{Comparison of analyses}
\begin{figure}
\centering
\includegraphics[width=1.\columnwidth]{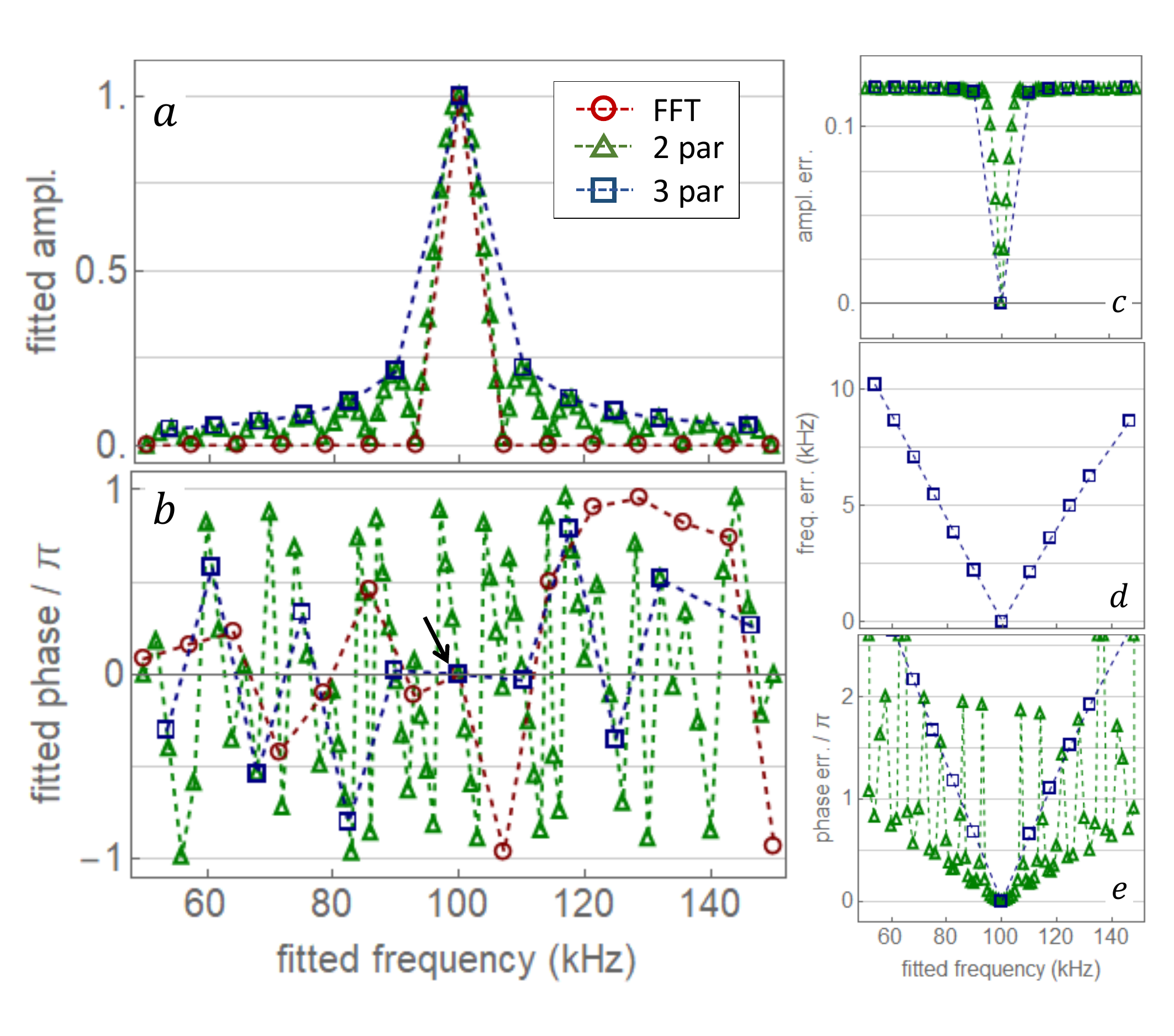}
\caption{\label{fig:cos}
\textbf{Comparison of all analysis methods for a pure cosine.}
\textbf{a,b.}
Results of the FFT (red circle), two-parameter (green triangle) and three-parameter fit (blue square) applied to a pure cosine function.
\textbf{c,d,e.}
$1\sigma$ fitting errors for the relevant analyses.
}
\end{figure}

As an illustrative example we apply all three analysis methods to a pure cosine with unity amplitude and zero phase: $\cos(2\pi\nu_p t)$ with $\nu_p=100$ kHz, with the same sample length and sampling frequency as our experimental data and as the fake signals used in the likelihood analysis.
The result of a Fourier transform is, of course, known analytically in this case (convolution of a $\delta$- with a sinc-function).
In Fig.~\ref{fig:cos}(a) one sees that the two-parameter fit is the closest to the analytic Fourier transform curve.
While the FFT captures the minima of $A(\nu)$, the three-parameter fit converges towards the local maxima.
This illustrates that the three-parameters fit method is the most robust because its SNR is the lowest.
In other words, it is the most sensitive measure for the strongest frequency components which are present in the signal and, as a consequence, it poses the most stringent test (see also likelihood analysis in the next section). 
The $\varphi(\nu)$ plot in Fig.~\ref{fig:cos}(b) shows that $\varphi^\star=0$ is correctly obtained at $\nu_p$ by all three methods; note though, that if the binning of the FFT would not have a point at this frequency, this information would be lost.
Therefore, either one of the two fitting methods is needed to correctly identify the phase.
Figs.~\ref{fig:cos}(c,d,e) also affirm that the fitting error is minimal for $\nu_p$ for all fitting parameters.
As a side remark, we mention that the FFT becomes more problematic if $\nu_p$ lies towards the center of a frequency bin.
It is well-known that a FFT analysis is best suited for large sample sizes -- a fact illustrated here.

\subsection{Likelihood analysis}
As the SNR of the experimental data is small the natural question that we have to answer is: what is the probability of finding a similar peak in random data (with the same standard deviation of noise) by applying our analyses.
As we have already mentioned, the most stringent test of this question is provided by the three-parameter fit, reflecting the reason why only this method is discussed in the main text.
It emphasizes the true strongest frequency components over the entire signal by searching for a local minimum of all three parameters.
If the frequency is fixed, as in two-parameter fit, more local minima will potentially be discovered in any type of noisy signal. 
We illustrate this idea by performing the following likelihood analysis.
  
After generating $1000$ fake signals all three analysis methods are conducted.
Each fake signal has the same length and sampling frequency as the real data and is drawn from a random Gaussian distribution with a standard deviation equal to the mean atom number standard deviation of the real data shown in Fig.~\ref{fig:numbVStime}(a) of the main text ($0.038$).
Then the analyses are conducted for each fake signal and we look for the events that show a SNR larger than that of the experiment (Table~\ref{tb:frequency}) for any dominant frequency over the entire frequency range $[50, 150]$ kHz.
For methods (i), (ii) and (iii) we find $1$, $1$ and $0$ such events respectively.
In addition, the two false positives are for different fake signals, i.e. there is no overlap (in contrast to the experimental data where all three analyses agree with each other).
If we lower the threshold to $\textrm{SNR}-\Delta(\textrm{SNR})$, thus allowing for more false events (however suppressed by a factor of $0.32$), $20$, $60$ and $1$ events are identified respectively (still no overlap).
It is now evident that the most stringent test for false events is provided by the three-parameters fit method.
Note, that to count the false positives for the FFT analysis we use the $1\sigma$ error of method (ii), because the FFT method does not yield an error estimation (see Table~\ref{tb:frequency}).
Narrowing the frequency range to the expected interval $[90,110]$ kHz reduces the number of false positives to $3$, $6$ and $0$ respectively.
If the threshold is lowered further to $\textrm{SNR}-2\Delta(\textrm{SNR})$ (suppression factor of $0.05$) $15$ overlapped fake signals are misinterpreted by all three methods.
Even in this case the overall probability to obtain a false positive is $7.5\times10^{-4}$.

\end{document}